\RequirePackage{ifpdf}
\ifpdf 
\documentclass[pdftex]{sigma}
\else
\documentclass{sigma}
\fi

\begin{document}

\allowdisplaybreaks
\renewcommand{\PaperNumber}{098}

\FirstPageHeading

\renewcommand{\thefootnote}{$\star$}

\ShortArticleName{Invariant Varieties of Periodic Points for the Discrete 
Euler Top}

\ArticleName{Invariant Varieties of Periodic Points\\ for the Discrete 
Euler Top\footnote{This paper is a contribution 
to the Vadim Kuznetsov Memorial Issue ``Integrable Systems and Related Topics''.
The full collection is available at 
\href{http://www.emis.de/journals/SIGMA/kuznetsov.html}{http://www.emis.de/journals/SIGMA/kuznetsov.html}}}

\Author{Satoru SAITO~$^\dag$ and Noriko SAITOH~$^\ddag$}
\AuthorNameForHeading{S. Saito and N. Saitoh}

\Address{$^\dag$~Hakusan 4-19-10, Midori-ku, Yokohama 226-0006, Japan}
\EmailD{\href{mailto:saito@phys.metro-u.ac.jp}{saito@phys.metro-u.ac.jp}} 

\Address{$^\ddag$~Applied Mathematics, Yokohama National University,\\
$\phantom{^\ddag}$~Hodogaya-ku, Yokohama 240-8501, Japan}
\EmailD{\href{mailto:nsaitoh@ynu.ac.jp}{nsaitoh@ynu.ac.jp}} 

\ArticleDates{Received October 28, 2006, in f\/inal form
December 16, 2006; Published online December 30, 2006}

\Abstract{The behaviour of periodic points of discrete Euler top is studied. 
We derive invariant varieties of periodic points explicitly. 
When the top is axially symmetric they are specif\/ied by some particular 
values of the angular velocity along the axis of symmetry, dif\/ferent for each period.}


\Classification{14J81; 39A11; 70E40} 

\begin{flushright}
\it To the memory of Professor Vadim B.~Kuznetsov
\end{flushright}

\section{Introduction}

The Kowalevski workshop on mathematical methods of regular dynamics was organized 
by Professor Vadim Kuznetsov in April 2000 at the University of Leeds \cite{Kowalevski workshop}. 
In his introductory talk about the Kowalevski top, Professor Kuznetzov \cite{Kuznetsov} had 
shown his strong interest on the subject and motivated the authors to work on classical tops. 

In our recent paper \cite{SS} we have studied the behaviour of periodic
 points of a rational map and found that they form a variety for each 
 period specif\/ied by invariants of the map if the map is integrable, 
 while they form a set of isolated points dependent on the invariants otherwise. 
 It is apparent that an application of our theorem to the problems of a classical 
 top is quite interesting and will be fruitful. We investigate the discrete Euler top, 
 in this article, to see how the invariant varieties of periodic points look like in 
 this particular example. In conclusion we will show that there is no periodic points 
 of period 2 and 4 if the top is not axially symmetric. In the case of period 3 we 
 derive explicitly an algebraic variety of dimension two as an invariant variety of periodic  points. 
 When the top is axially symmetric, the angular velocity of a periodic map along the symmetry 
 axis is quantized to some special values determined by the period and the shape of the top. 
 The other components of the angular velocity are free, thus form an invariant variety of 
 periodic points separately for each period.

To start with let us brief\/ly review our theorem of~\cite{SS}. 
We consider a rational map on $\hat{\mathbf C}^d$, where $\hat{\mathbf C}=\{{\mathbf C},\infty\}$,
\begin{gather}
{\mathbf x}=(x_1,x_2,\dots ,x_d)\quad \rightarrow\quad
{\mathbf X}=(X_1,X_2,\dots ,X_d)=:{\mathbf X}^{(1)}.
\label{x->X}
\end{gather}
We are interested in the behaviour of the sequence: $
{\mathbf x}\rightarrow {\mathbf X}^{(1)}\rightarrow{\mathbf X}^{(2)}\rightarrow\cdots$. 
In particular we pay attention to the behaviour of periodic points of rational maps. 
If the map is nonintegrable we shall f\/ind a set of isolated points with fractal structure 
as a higher dimensional counterpart of the Julia set. Our question in \cite{SS} 
was what object appears when the map is integrable.

We assume that the map has $p\ (\ge 0)$ invariants 
$H_1({\mathbf x}),H_2({\mathbf x}),\dots  H_p({\mathbf x})$. 
If $h_1,h_2,\dots ,h_p$ are the values of the invariants given by 
the initial values of the map, the orbit of the map is constrained on 
the $d-p$ dimensional variety determined by the conditions $H_i({\mathbf x})=h_i$,
$i=1,2,\dots ,p$, which we denote by $V(h)$. Now let us consider the periodicity 
conditions ${\mathbf X}^{(n)}={\mathbf x}$ of period $n$. 
We can eliminate $p$ variables out of ${\mathbf x}$ and only $d-p$ 
periodicity conditions remain. If they are independent, we obtain 
a set of isolated periodic points on $V(h)$ in general. 

It may happen, however, that some of the $d-p$ conditions impose some 
relations on $h_1,h_2,\dots$, $h_p$ instead of f\/ixing all of the $d-p$ variables. 
If $m$ is the number of the conditions which determine the values of the variables, 
$d-p-m$ variables are left free and the periodic points of period $n$ 
form a subvariety of dimension $d-p-m$ on $V(h)$, instead of a set 
of isolated points. In this case we say that the periodicity conditions 
are `correlated'. If $l$ is the number of the periodicity conditions 
which relate the invariants, $p-l$ invariants remain undetermined. 
This means that the periodic points of period $n$ form a $d-l-m$ 
dimensional subvariety in $\hat {\mathbf C}^d$. We have proven in \cite{SS} the following lemma:

\begin{lemma}[\cite{SS}]\label{lemma1}
A set of correlated periodicity conditions satisfying 
$
\min\{p,\ d-p\}\ge l+m
$
and a set of uncorrelated periodicity conditions of a dif\/ferent period do not exist in one map simultaneously.
\end{lemma}

When $m=0$ the periodicity conditions determine none of the variables but 
impose $l$ relations among the invariants. In this particular case all points of $V(h)$ 
are the points of period $n$, while the variety $V(h)$ itself is constrained by the 
relations among the invariants. We call the periodicity conditions are `fully correlated' 
in this case. The periodic points form a subvariety of dimension $d-l$ in 
the space $\hat{\mathbf C}^d$, which we call `an invariant variety of periodic points'. 
Every point of this variety can be an initial point of the $n$ period map, whose orbit 
stays on it. Since the condition $\min\{p,\ d-p\}\ge l+m$ is automatically satisf\/ied, 
our theorem follows to the Lemma~\ref{lemma1} immediately:

\begin{theorem}[\cite{SS}]\label{theorem1}
    If there is an invariant variety of periodic points of some period, 
    there is no set of isolated periodic points of other period in the map.
\end{theorem}

The Theorem \ref{theorem1} doesn't tell us directly whether the map is integrable 
or nonintegrable. There is, however, some evidence to believe that the periodic 
points of a nonintegrable map, if they exist, form a fractal set of isolated points. 
Therefore it is reasonable to adopt the following proposition as our working hypothesis:

\smallskip

{\it If a map is nonintegrable, there is a set of uncorrelated periodicity conditions of some period}.

\smallskip

The Julia set, which is the source of chaotic orbits, is a subset of the closure of all 
isolated periodic points. We emphasize that our hypothesis does not require that all 
of the periodicity conditions of a nonintegrable map are uncorrelated but requires 
only one at least. On the other hand our theorem shows that the existence of an 
invariant variety of periodic points excludes a set of isolated periodic points 
in the map and vice versa. This means that if a set of periodic points of some 
period forms an invariant variety there is no Julia set, thus suggesting the following statement:

\medskip

\noindent
{\bf Conjecture (\cite{SS})}.
{\it 
If there is an invariant variety of periodic points of some period, the map is integrable.}

\medskip

Note that this does not exclude possibilities that some integrable maps do 
not have an invariant variety of periodic points. For example an integrable 
map with no invariant does not have an invariant variety. 
On the other hand there are $d$ dimensional maps which have $d-1$ 
invariants but reduce to the logistic map after elimination of $d-1$ 
variables by using the invariants. Therefore the situation is quite dif\/ferent 
from the continuous time Hamiltonian f\/low, whose integrability is guaranteed 
by the Liouville theorem if there are suf\/f\/icient number of invariants.

In order to support our conjecture we have studied in \cite{SS} various maps, 
such as the QRT maps, the Lotka--Volterra maps and the Toda maps, 
and found that all periodic points form invariant varieties of periodic points 
if the map is integrable and periodic points exist, while no such property 
has been found otherwise. We also studied the $q$-Painlev\'e maps which 
are integrable but not volume preserving in general. We found invariant 
varieties only when the parameters are restricted so that the maps have suf\/f\/icient number of invariants.

We introduce the discrete Euler top in Section~2. To f\/ind an invariant variety of periodic points for 
the Euler top the general scheme developed in \cite{SS} is applied in Sections~3 and 4. 
In the f\/inal section we study explicitly the nature of the invariant surfaces of an axially symmetric Euler top.

\section{Discrete Euler top}

When the time is continuous, the equation of motion for the Euler top is given by
\[
I_1{d\omega_1\over dt}=(I_2-I_3)\omega_2\omega_3,\qquad
I_2{d\omega_2\over dt}=(I_3-I_1)\omega_3\omega_1,\qquad
I_3{d\omega_3\over dt}=(I_1-I_2)\omega_1\omega_2,
\]
where $(\omega_1, \omega_2, \omega_3)$ are the angular velocity in the body f\/ixed frame 
and $(I_1,I_2,I_3)$ are the corresponding moments of inertia. 
The system has two invariants, ${1\over 2}(I_1\omega_1^2+I_2\omega_2^2+I_3\omega_3^2)$ and 
$I_1^2\omega_1^2+I_2^2\omega_2^2+I_3^2\omega_3^2$, corresponding to the total 
kinetic energy and the square of angular momentum, hence is integrable.

A discretization of the Euler equation, which preserves integrability, was f\/irst obtained by 
Bobenko et al.~in~\cite{BLS}, and then discussed by other authors~\cite{HK,HK+, F, MW}. 
We adopt here an explicit version of the discretization proposed by Hirota et al.~\cite{HK,HK+}. 
After the discretization we write the angular velocity as $(x_1,x_2,x_3)$ instead of 
$(\omega_1,\omega_2,\omega_3)$ and consider the map $(x_1,x_2,x_3)\rightarrow (X_1,X_2,X_3)$ def\/ined by
\begin{gather}
I_1(X_1-x_1)={\delta\over 2}(I_2-I_3)(X_2x_3+x_2X_3),\nonumber\\
I_2(X_2-x_2)={\delta\over 2}(I_3-I_1)(X_3x_1+x_3X_1),
\label{Euler map}\\
I_3(X_3-x_3)={\delta\over 2}(I_1-I_2)(X_1x_2+x_1X_2).\nonumber
\end{gather}
The continuous limit corresponds to $\delta\rightarrow 0$.

Solving the equation (\ref{Euler map}) for $(X_1,X_2,X_3)$ we f\/ind
\begin{gather}
X_1={x_1(1-\alpha_2\alpha_3 x_1^2+\alpha_3\alpha_1x_2^2+\alpha_1\alpha_2x_3^2)+2\alpha_1x_2x_3\over
1-2\alpha_1\alpha_2\alpha_3x_1x_2x_3-\alpha_2\alpha_3 x_1^2-\alpha_3\alpha_1x_2^2-\alpha_1\alpha_2x_3^2},\nonumber\\
X_2={x_2(1+\alpha_2\alpha_3 x_1^2-\alpha_3\alpha_1x_2^2+\alpha_1\alpha_2x_3^2)+2\alpha_2x_3x_1\over
1-2\alpha_1\alpha_2\alpha_3x_1x_2x_3-\alpha_2\alpha_3 x_1^2-\alpha_3\alpha_1x_2^2-\alpha_1\alpha_2x_3^2},
\label{X=}\\
X_3={x_3(1+\alpha_2\alpha_3 x_1^2+\alpha_3\alpha_1x_2^2-\alpha_1\alpha_2x_3^2)+2\alpha_3x_1x_2\over
1-2\alpha_1\alpha_2\alpha_3x_1x_2x_3-\alpha_2\alpha_3 x_1^2-\alpha_3\alpha_1x_2^2-\alpha_1\alpha_2x_3^2},\nonumber
\end{gather}
where we used the notations
\[
\alpha_1=\delta{I_2-I_3\over 2I_1},\qquad 
\alpha_2=\delta{I_3-I_1\over 2I_2},\qquad 
\alpha_3=\delta{I_1-I_2\over 2I_3}.
\]

We notice that, when the top is axially symmetric, the map (\ref{X=}) 
is nothing but a two dimensional linear transformation. For example if we assume $I_2=I_3$, the map becomes
\begin{gather}
X_1=x_1,\nonumber\\
X_2=x_2\cos\Omega + x_3\sin\Omega,
\label{axially symmetric top}\\
X_3=x_3\cos\Omega -x_2\sin\Omega ,\nonumber
\end{gather}
where
\[
\cos\Omega={4I_2^2-(I_2-I_1)^2x_1^2\over 4I_2^2+(I_2-I_1)^2x_1^2}.
\]
Therefore we discuss the axially symmetric top separately from the generic case in Section~5.

The invariants of the map (\ref{X=}) are
\begin{gather}
H_1={I_1x_1^2+I_2x_2^2+I_3x_3^2\over 1-\alpha_2\alpha_3 x_1^2},\qquad
H_2={I_1^2x_1^2+I_2^2x_2^2+I_3^2x_3^2\over 1-\alpha_2\alpha_3 x_1^2}, 
\label{invariants}
\end{gather}
as it will be checked by a direct substitution of (\ref{X=}). They coincide with the invariants of 
the continuous case in the limit $\delta\rightarrow 0$. We f\/ix the value of $\delta$ at 1 hereafter, 
since it is irrelevant in the following discussions.

If we denote by $x$ one of the three variables $(x_1,x_2,x_3)$, the elimination of other two 
variables from the map (\ref{Euler map}) yields
\begin{gather}
S(X,x;{\bf q})=0
\label{S=0}
\end{gather}
with
\begin{gather}
S(X,x;{\bf q}):=aX^2x^2+bXx(X+x)+c(X-x)^2+dXx+e(X+x)+f.
\label{S(X,x,q)}
\end{gather}
The parameters ${\mathbf q}=({\mathbf a},{\mathbf b},{\mathbf c},{\mathbf d},{\mathbf e},{\mathbf f})$ are given by
\begin{gather}
\left(\begin{array}{c}
a_1\\
b_1\\
c_1\\
d_1\\
e_1\\
f_1\\
\end{array}\right)
=\left(\begin{array}{l}
-4\alpha_2\alpha_3(A_0-A_2)(A_0+A_3)\\
0\\
(A_0-A_2+A_3)^2\\
4\Big(A_1^2-A_2(A_0+A_3)+A_3(A_0-A_2)\Big)\\
0\\
(4/\alpha_2\alpha_3)A_2A_3\\
\end{array}\right),\nonumber\\
\left(\begin{array}{c}
a_2\\
b_2\\
c_2\\
d_2\\
e_2\\
f_2\\
\end{array}\right)
=\left(\begin{array}{l}
-4\alpha_3\alpha_1A_0(A_0-A_2)\\
0\\
(A_0-A_2-A_3)^2\\
4\Big(A_2^2-A_0(A_2+A_3)-A_3(A_0-A_2)\Big)\\
0\\
(4/\alpha_3\alpha_1)A_3A_1\\
\end{array}\right),
\label{q of Euler top}
\\
\left(\begin{array}{c}
a_3\\
b_3\\
c_3\\
d_3\\
e_3\\
f_3\\
\end{array}\right)
=\left(\begin{array}{l}
-4\alpha_1\alpha_2A_0(A_0+A_3)\\
0\\
(A_0+A_2+A_3)^2\\
4\Big(A_3^2+A_0(A_2+A_3)+A_2(A_0+A_3)\Big)\\
0\\
(4/\alpha_1\alpha_2)A_1A_2\\
\end{array}\right)\nonumber
\end{gather}
corresponding, respectively, to $x=x_1, x_2, x_3$. Here we introduced the notations
\begin{gather*}
A_0=4I_1I_2I_3,\qquad
A_1=(I_2-I_3)(I_1H_1-H_2),\qquad
A_2=(I_3-I_1)(I_2H_1-H_2),\\
A_3=(I_1-I_2)(I_3H_1-H_2),\qquad \big( A_1+A_2+A_3=0\big).
\end{gather*}

One might wonder that the map $x\rightarrow X$ def\/ined by (\ref{S=0}) 
does not determine an image of the map uniquely. Since the function $S(X,x;{\mathbf q})$ 
of (\ref{S(X,x,q)}) is symmetric under the exchange of the variables $x$ and $X$,
we see that the two solutions of (\ref{S=0}) correspond to the forward 
and the backward maps, which we denote $X^{(1)}$ and $X^{(-1)}$. 
If we apply the map to $X^{(1)}$ we should get~$x$ and $X^{(2)}$. 
In this way we shall obtain a chain of images of the map into two directions:
\[
\cdots \longleftarrow X^{(-2)}\longleftarrow X^{(-1)}\longleftarrow x\longrightarrow X^{(1)}\longrightarrow X^{(2)}\longrightarrow \cdots .
\]
The number of branches of the map does not increase, but remains two, at every step of the map. 
Therefore a map of the form (\ref{S=0}) is well def\/ined in general.
\section{Iteration of the map}

We studied in \cite{SS} the map def\/ined by (\ref{S=0}) and called it a `biquadratic map'.
 An iteration of the map yields the biquadratic map again but with new parameters ${\mathbf q}^{(2)}$:
\begin{gather}
a^{(2)}:= (ae-cb)^2-(ad-2ac-b^2)(be-cd+2c^2),\nonumber\\
b^{(2)}:= (ae-cb)(2af-be+cd-4c^2)-(ad-2ac-b^2)(bf-ce),\nonumber\\
c^{(2)}:=(af-c^2)^2-(ae-bc)(bf-ce),\nonumber\\
d^{(2)}:=4(af-c^2)^2-2(ae-bc)(bf-ce)-(be-cd+2c^2)^2\label{a_2,...,f_2}\\
\phantom{d^{(2)}:=}{}-(ad-2ac-b^2)(df-2cf-e^2),\nonumber\\
e^{(2)}:= (fb-ce)(2af-be+cd-4c^2)-(fd-2fc-e^2)(ea-cb),\nonumber\\
f^{(2)}:= (fb-ce)^2-(fd-2fc-e^2)(be-cd+2c^2).\nonumber
\end{gather}

If we repeat the map further we obtain a series of biquadratic 
maps whose parameters can be determined iteratively from the previous ones as follows:
\begin{gather}
a^{(n+1)}={1\over a^{(n-1)}}\big((a_\wedge c)_n^2-(a_\wedge b)_n(b_\wedge c)_n\big),\nonumber\\
b^{(n+1)}={1\over a^{(n-1)}}\Bigg({b^{(n-1)}\over a^{(n-1)}}\big((a_\wedge b)_n(b_\wedge c)_n-(a_\wedge c)_n^2\big)
+(a_\wedge c)_n\big((a_\wedge e)_n+2(b_\wedge c)_n\big)\nonumber\\
\phantom{b^{(n+1)}=}{}- {1\over 2}\big((a_\wedge b)_n(b_\wedge e)_n-(a_\wedge b)_n
(c_\wedge d)_n+(a_\wedge d)_n(b_\wedge c)_n\big)\Bigg),\nonumber\\
c^{(n+1)}={1\over 2c^{(n-1)}}\Big((ce^{(n)}-bf^{(n)})(ae^{(n)}-bc^{(n)})+(cb^{(n)}
-ea^{(n)})(fb^{(n)}-ec^{(n)})\nonumber\\
\phantom{c^{(n+1)}=}{} +(af^{(n)}-cc^{(n)})^2+(fa^{(n)}-cc^{(n)})^2\Big),\label{n+1th parameters}\\
d^{(n+1)} = {1\over d^{(n-1)}}\Big(-f^{(n-1)}a^{(n+1)}-a^{(n-1)}f^{(n+1)}-4b^{(n-1)}e^{(n+1)}
-4e^{(n-1)}b^{(n+1)}+(a_\wedge f)_n^2\nonumber\\
\phantom{d^{(n+1)} =}{}+(c_\wedge d)_n^2-(a_\wedge b)_n(e_\wedge f )_n-(b_\wedge c)_n(c_\wedge e)_n
+(a_\wedge d)_n(d_\wedge f)_n+2(b_\wedge e)_n(a_\wedge f)_n\nonumber\\
\phantom{d^{(n+1)} =}{}-
\big(3(c_\wedge e)_n-(b_\wedge f)_n-(d_\wedge e)_n\big)
\big(3(b_\wedge c)_n-(a_\wedge e)_n-(b_\wedge d)_n\big)\nonumber\\
\phantom{d^{(n+1)} =}{}+2\big((a_\wedge d)_n-(a_\wedge c)_n\big)\big((c_\wedge f)_n-(d_\wedge f)_n\big)+
2\big((b_\wedge c)_n+(a_\wedge e)_n\big)\big((b_\wedge f)_n+(c_\wedge e)_n\big)\!\Big),\!\!\nonumber
\\
e^{(n+1)}={1\over f^{(n-1)}}\Bigg({e^{(n-1)}\over f^{(n-1)}}\big((f_\wedge e)_n(e_\wedge c)_n-(f_\wedge c)_n^2\big)
+(f_\wedge c)_n\big((f_\wedge b)_n+2(e_\wedge c)_n\big)\nonumber\\
\phantom{e^{(n+1)}=}{}
-{1\over 2}\big((f_\wedge e)_n(e_\wedge b)_n-(f_\wedge e)_n(c_\wedge d)_n
+(f_\wedge d)_n(e_\wedge c)_n\big)\Bigg),\nonumber\\
f^{(n+1)}=
{1\over f^{(n-1)}}\big((f_\wedge c)_n^2-(f_\wedge e)_n(e_\wedge c)_n\big),\nonumber
\end{gather}
where we used the notation $(g_\wedge g')_n=g{g'}^{(n)}-g'g^{(n)}$.

Despite the complicated expression of the relation (\ref{n+1th parameters}), we observe 
a special dependence on the $n$th parameters ${\mathbf q}^{(n)}$. 
Besides $c^{(n+1)}$, the dependence of the $(n+1)$th parameters 
on the $n$th ones is always in the form $(g_\wedge g')_n=g{g'}^{(n)}-g'g^{(n)}$. 
They all vanish simultaneously when the parameters ${\mathbf q}^{(n)}$ are `fully correlated', 
that is, if there exists a function $\gamma^{(n+1)}(\mathbf{q})$ such that
\begin{gather}
{\mathbf q}^{(n)}=\epsilon{\mathbf q}+\gamma^{(n+1)}({\mathbf q})\hat{\mathbf q}^{(n)},
\label{a_n=a+gamma a}
\end{gather}
where $\epsilon$ is an arbitrary constant. In fact we obtain, after some manipulation,
\begin{gather*}
(a_\wedge b)_2=(af-eb-3c^2+cd)(2a^2e-abd+b^3),\\
(a_\wedge c)_2=(af-eb-3c^2+cd)(a^2f+ac^2-acd+b^2c),\\
(b_\wedge c)_2=(af-eb-3c^2+cd)(2ace-abf-bc^2),\\
\cdots\cdots\cdots\cdots\cdots\cdots\cdots\cdots\cdots\cdots\cdots\cdots\cdots\cdots\cdots\cdots\\
(e_\wedge f)_2=(af-eb-3c^2+cd)(edf-e^3-2bf^2),
\end{gather*}
from which we f\/ind $\gamma^{(3)}(\mathbf{q})$:
\begin{gather}
\gamma^{(3)}({\mathbf q})=af-be-3c^2+cd.
\label{gamma3general}
\end{gather}
The formula (\ref{n+1th parameters}) enables us to f\/ind a series of $\gamma^{(n)}(\mathbf{q})$ 
iteratively, as follows:
\begin{gather}
\gamma^{(4)}(\mathbf{q})=2acf-adf+b^2f+ae^2-2c^3+c^2d-2bce,\label{gamma4general}\\
\gamma^{(5)}(\mathbf{q})=
a^3f^3+\Big(-cf^2d+2cfe^2+fde^2-3ebf^2-e^4-c^2f^2\Big)a^2\nonumber\\
\phantom{\gamma^{(5)}(\mathbf{q})=}{}+\Big(-13c^4f+18c^3fd+de^3b+2cf^2b^2+7dc^2e^2-ce^2d^2-2ce^3b\nonumber\\
\phantom{\gamma^{(5)}(\mathbf{q})=}{}+2c^2feb-7fd^2c^2-14c^3e^2+cd^3f+fb^2e^2+f^2db^2-ebd^2f\Big)a\label{gamma_5general}\\
\phantom{\gamma^{(5)}(\mathbf{q})=}{}- cd^2b^2f-b^3e^3-4c^3deb+cdb^2e^2+13ec^4b-f^2b^4+7fb^2c^2d\nonumber\\
\phantom{\gamma^{(5)}(\mathbf{q})=}{}+ c^4d^2-5c^5d+5c^6-2fb^3ec-e^2c^2b^2+eb^3df-14fb^2c^3,
\nonumber
\end{gather}
and so on. 

When (\ref{a_n=a+gamma a}) holds, the equation $S(Q,x;{\mathbf q}_{n+1})=0$ can be written as
\begin{gather}
c^{(n+1)}(Q-x)^2+(\gamma^{(n+1)}({\mathbf q}))^2K_{n+1}(Q,x)=0,\qquad n=2,3,4,\dots. 
\label{c_n(Q-x)^2+gamma_nK_n(Q,x)=0}
\end{gather}
Here 
\begin{gather*}
K_{n+1}(Q,x)=\hat a^{(n+1)}Q^2x^2+\hat b^{(n+1)}(Q+x)Qx+\hat d^{(n+1)}Qx+\hat e^{(n+1)}(Q+x)+\hat f^{(n+1)},
\end{gather*}
and $\hat a^{(n+1)}$, for instance, is obtained from $a^{(n+1)}$ 
simply replacing $(g_\wedge g')_n$ by $(\hat g_\wedge{\hat g}')_n$. 
If $Q$ is a~point of period $n+1$, the f\/irst term of (\ref{c_n(Q-x)^2+gamma_nK_n(Q,x)=0}) 
vanishes. Hence the periodicity condition requires for the second 
term to vanish. This is certainly satisf\/ied for arbitrary $x$ if $\gamma^{(n+1)}(\mathbf{q})=0$, 
namely when the periodicity conditions for the parameters ${\mathbf q}^{(n)}$ are fully correlated. 
The other possible solutions obtained by solving $K_{n+1}(x,x)=0$ do not correspond to the points 
of period $n+1$, but represent the f\/ixed points or the points of periods which divide $n+1$.

\section{Invariant varieties of periodic points\\ for the discrete Euler top}

We are ready to study the periodicity conditions for the discrete Euler top. Throughout 
this section we will not consider axially symmetric cases, which we discuss in the next section. 
The direct calculation of the periodicity conditions $X_j^{(n)}=x_j$ is not easy to carry out 
by a small computer. Therefore we use the method we developed in the previous section. 
Before starting, however, let us f\/irst search the f\/ixed points of the map. 
If we remember that the variables $(x_1,x_2,x_3)$ are the discrete analog of the angular 
velocity $(\omega_1,\omega_2,\omega_3)$, a f\/ixed point of the map corresponds to 
the motion of the top which does not change the angular velocity in all directions of the body f\/ixed frame. 

Needless to say the f\/ixed points are nothing to do with the invariants of the map. 
To f\/ind them we go back to the map (\ref{Euler map}) and see immediately that they are
\[
{\rm fixed\ points:}\quad\{{\mathbf x}\; | \;x_2=x_3=0\, \cup\, x_3=x_1=0\, \cup \, x_1=x_2=0\}.
\]
This result shows that a f\/ixed point is realized as a steady rotation along one of the three axes. 
The value of the angular velocity along the direction is arbitrary, while the angular velocities 
are zero along the other two directions.

The method we developed in the previous section enables us to get information of 
the periodicity conditions of period greater than three. The case of period 2 must be considered separately. 
From the general expression (\ref{a_2,...,f_2})
 the parameters in $S(X^{(2)},x;{\mathbf q}_2)$ are not proportional to a 
 common factor. But if we substitute (\ref{q of Euler top}) into (\ref{a_2,...,f_2}), 
 we f\/ind that they have the following form
\begin{gather*}
a^{(2)}=(\gamma^{(2)}({\mathbf q}))^2\hat a^{(2)},\qquad b^{(2)}=0,\qquad d^{(2)}
=(\gamma^{(2)}({\mathbf q}))^2\hat d^{(2)},\\
 e^{(2)}=0,\qquad f^{(2)}=(\gamma^{(2)}({\mathbf q}))^2\hat f^{(2)}
\end{gather*}
with
\begin{gather}
\gamma_1^{(2)}({\mathbf q})=A_0-A_2+A_3,\qquad
\gamma_2^{(2)}({\mathbf q})=A_0-A_2-A_3,\qquad
\gamma_3^{(2)}({\mathbf q})=A_0+A_2+A_3,
\label{gamma2}
\end{gather}
corresponding to $x=x_1,\ x_2,\ x_3$, respectively. For the higher periods we can apply the 
formulae~(\ref{gamma3general}) and~(\ref{gamma4general}) to obtain
\begin{gather}
\gamma_1^{(3)}=\big(A_1^2+2A_0(A_2-A_3)-3A_0^2\big)\big((A_1+A_0)^2+4A_0A_3\big),\nonumber\\
\gamma_2^{(3)}=\big(A_1^2+2A_0(A_2-A_3)-3A_0^2\big)\big((A_1+A_0)^2+4A_3A_1\big),
\label{gamma3}\\
\gamma_3^{(3)}=\big(A_1^2+2A_0(A_2-A_3)-3A_0^2\big)\big((A_1-A_0)^2+4A_1A_2\big),\nonumber
\\
\gamma_1^{(4)}=2(A_1-A_0)(A_0-A_2-A_3)\big((A_1-A_0)^4-8A_2(A_0+A_3)(A_1^2+A_0^2)\big),\nonumber\\
\gamma_2^{(4)}=2(A_0-A_1)(A_0-A_2+A_3)\big((A_1+A_0)^4+16A_0A_1A_3(A_0-A_2)\big),
\label{gamma4}\\
\gamma_3^{(4)}=2(A_0+A_1)(A_0-A_2+A_3)\big((A_1-A_0)^4+16A_0A_1A_2(A_0+A_3)\big),\nonumber\\
\vdots\nonumber
\end{gather}

From the expressions (\ref{gamma2}), (\ref{gamma3}), (\ref{gamma4}) 
it is clear that the periodicity conditions do not determine points but 
impose relations among the invariants of the map. This owes to the fact that the 
initial parameters ${\mathbf q}$ are dependent on the invariants of the map, as 
we see in (\ref{q of Euler top}). After iteration of the map $n$ times the new 
parameters ${\mathbf q}^{(n)}$ are also dependent on the invariants. Therefore 
the periodicity condition $\gamma^{(n)}({\mathbf q})=0$ imposes relations among the invariants.

The periodicity conditions of period $n$ are satisf\/ied when $\gamma_1^{(n)}=0$,
$\gamma_2^{(n)}=0$, $\gamma_3^{(n)}=0$ are satisf\/ied 
simultaneously. By an inspection of (\ref{gamma2}), (\ref{gamma3}), (\ref{gamma4}) 
we notice that the conditions are satisf\/ied by the single condition
\begin{gather}
A_1^2+2A_0(A_2-A_3)-3A_0^2=0
\label{rarara}
\end{gather}
in the case of $n=3$. All other cases impose further relations among the invariants.

In order to f\/ind where the periodic points are in the $(x_1,x_2,x_3)$ space, 
we simply substitute the formulae (\ref{invariants}) into the conditions 
$\gamma^{(n)}({\mathbf q})=0$. In the case of (\ref{rarara}) we obtain
\begin{gather}
v^{(3)}=\left\{{\mathbf x}\; \big|\;(1+\xi_1+\xi_2+\xi_3)^2-4(1+\xi_1\xi_2+\xi_2\xi_3+\xi_3\xi_1)=0\right\},
\label{rururu}
\end{gather}
in terms of the new variables
\[
\xi_1={(I_3-I_1)(I_1-I_2)\over 4I_2I_3}x_1^2,\qquad
\xi_2={(I_1-I_2)(I_2-I_3)\over 4I_3I_1}x_2^2,\qquad
\xi_3={(I_2-I_3)(I_3-I_1)\over 4I_1I_2}x_3^2.
\]
The set of points satisfying (\ref{rururu}) form a variety 
of periodic points of period 3 in the space of $(x_1,x_2,x_3)$. 
Every point on this variety is a point of period~3. We called this 
type of variety `an invariant variety of periodic points', because 
it is determined uniquely by the invariants of the map alone. The 
dimension of the variety is two, which is the number of the invariants. 
In the case of (\ref{rururu}) the invariant variety is an algebraic variety of degree 4, 
symmetric in the three variables $x_1$, $x_2$, $x_3$.

Now let us pause a while. The discrete Euler top (\ref{Euler map}) 
has been known being satisf\/ied by elliptic functions as special solutions. 
The map generates an elliptic curve. This curve, however, is not the invariant 
variety of periodic points in our consideration, since the map is not controlled, 
in general, by the periodicity conditions of some f\/ixed period. In fact the 
invariant variety of (\ref{rururu}) is not a curve but a surface. Once an 
initial point is chosen on the surface, the orbit stays on it before it 
returns to the initial point. The invariant variety (\ref{rururu}) tells 
us where the map of period $3$ should start. Every point on (\ref{rururu}) 
is a candidate of the period $3$ map. The elliptic curve is embedded in 
this invariant variety as a set of $3$ points, if the initial point is on it. 
We can view this variety as a subspace of the set of all elliptic curves, 
which are restricted to $3$ periodic motion. It is a highly nontrivial 
observation that the intersections form a surface characterized by certain specif\/ic 
relations among the invariants of the map alone. The existence of such an variety in 
any integrable map has not been known, to our knowledge, in the literature. The claim 
of our conjecture is that if there exists an invariant variety of some period, the map 
is guaranteed being integrable. Therefore the existence of the surface (\ref{rururu}) of 
period 3 is suf\/f\/icient to guarantee the integrability of the discrete Euler top. This is 
true irrespective whether some solutions are known or not known explicitly. 

To see other conditions, let us present all expressions of (\ref{gamma2}), 
(\ref{gamma3}), (\ref{gamma4}) after the substitution of (\ref{invariants}):
\begin{gather}
\gamma_1^{(2)}=(1+\xi_1-\xi_2-\xi_3),\quad
\gamma_2^{(2)}=(1-\xi_1+\xi_2-\xi_3),\quad
\gamma_3^{(2)}=(1-\xi_1-\xi_2+\xi_3),
\label{gamma2xi}
\\
\gamma_1^{(3)}=
\big((1+\xi_1+\xi_2+\xi_3)^2-4(1+\xi_1\xi_2+\xi_2\xi_3+\xi_3\xi_1)\big)\nonumber\\
\phantom{\gamma_1^{(3)}=}{} \times\big((1+\xi_1-\xi_2-\xi_3)^2
-4(\xi^2_1-\xi_1\xi_2+\xi_2\xi_3-\xi_3\xi_1)\big),\nonumber\\
\gamma_2^{(3)}=
\big((1+\xi_1+\xi_2+\xi_3)^2-4(1+\xi_1\xi_2+\xi_2\xi_3+\xi_3\xi_1)\big)\label{gamm3xi}\\
\phantom{\gamma_2^{(3)}=}{}\times\big((1-\xi_1+\xi_2-\xi_3)^2-4(\xi^2_2-\xi_1\xi_2-\xi_2\xi_3+\xi_3\xi_1)\big),\nonumber\\
\gamma_3^{(3)}=
\big((1+\xi_1+\xi_2+\xi_3)^2-4(1+\xi_1\xi_2+\xi_2\xi_3+\xi_3\xi_1)\big)\nonumber\\
\phantom{\gamma_3^{(3)}=}{}\times\big((1-\xi_1-\xi_2+\xi_3)^2-4(\xi^2_3+\xi_1\xi_2-\xi_2\xi_3-\xi_3\xi_1)\big),\nonumber
\\
\gamma_1^{(4)}=\big((1-\xi_1)^2-(\xi_2-\xi_3)^2\big)
\big((\xi_1-1)^2\big(2(1+\xi_1-\xi_2-\xi_3)^2-(\xi_1-1)^2\big)\nonumber\\
\phantom{\gamma_1^{(4)}=}{}+(\xi_2-\xi_3)^2\big((2+2\xi_1-\xi_2-\xi_3)^2+4\xi_3\xi_2-8\xi_1\big)\big),\nonumber\\
\gamma_2^{(4)}=\big((1-\xi_2)^2-(\xi_3-\xi_1)^2\Big)\Big((\xi_2-1)^2\big(2(1-\xi_1+\xi_2-\xi_3)^2
-(\xi_2-1)^2\big)\label{gamma4xi}\\
\phantom{\gamma_2^{(4)}=}{} +(\xi_3-\xi_1)^2\big((2+2\xi_2-\xi_3-\xi_1)^2+4\xi_1\xi_3-8\xi_2\big)\big),\nonumber\\
\gamma_3^{(4)}=\big((1-\xi_3)^2-(\xi_1-\xi_2)^2\big)\big((\xi_3-1)^2\big(2(1-\xi_1-\xi_2+\xi_3)^2-(\xi_3-1)^2\big)
\nonumber\\
\phantom{\gamma_3^{(4)}=}{} +(\xi_1-\xi_2)^2\big((2+2\xi_3-\xi_1-\xi_2)^2+4\xi_1\xi_2-8\xi_3\big)\big).
\nonumber
\end{gather}

The conditions (\ref{gamma2xi}) for the period 2 impose
\begin{gather}
\xi_1=\xi_2=\xi_3=1.
\label{xi_1=xi_2=xi_3=1}
\end{gather}
The second factors of $\gamma^{(3)}_1$,
$\gamma^{(3)}_2$, $\gamma^{(3)}_3$ vanish simultaneously only when the point $(x_1,x_2,x_3)$ 
is on the lines def\/ined by
\begin{gather}
\{{\mathbf x}\,|\, \xi_1=1\; \cap\;  \xi_2=\xi_3\}\; \cup\; 
\{{\mathbf x}\,|\, \xi_2=1\; \cap\  \xi_3=\xi_1\}\; \cup\; 
\{{\mathbf x}\,|\, \xi_3=1\; \cap\;  \xi_1=\xi_2\}.
\label{three lines}
\end{gather}
After some manipulation we f\/ind that $\gamma^{(4)}_1$,
$\gamma^{(4)}_2$, $\gamma^{(4)}_3$ also vanish simultaneously if\/f the point is on these lines. 
Therefore every periodic point of period 2, 3 and 4 are on the lines of (\ref{three lines}), 
if it is not on the invariant variety (\ref{rururu}) of period 3. We now notice that the 
points on these lines (\ref{three lines}) vanish the denominator of the map (\ref{X=}), 
or equivalently the function $(1-\xi_1-\xi_2-\xi_3)^2-4\xi_1\xi_2\xi_3$.

From this observation we are convinced that the Euler top has no periodic 
point of period 2 and 4 as long as the top is not axially symmetric, 
whereas the periodic points of period 3 form the invariant variety $v^{(3)}$ of~(\ref{rururu}).

\section{Axially symmetric top}

By studying the discrete Euler top we have found an invariant variety 
of periodic points in the case of period 3. If we adopt our conjecture in Section~1, 
this means that the system is integrable. We also found that there is 
no periodic point of period 2 and 4 if the top is not axially symmetric. 
Our method enables us to search the periodic points of larger period. 
Instead of carrying out further the cumbersome algebraic analysis, however, 
we conclude this paper by studying the cases of symmetric top.

If the top is totally symmetric, i.e., $I_1=I_2=I_3$, 
the equations (\ref{X=}) show that $(x_1,x_2,x_3)$ remain constants. 
This is a top which never changes its angular velocity in all directions. 
When the top is axially symmetric, such as $I_2=I_3$, the motion is governed 
by (\ref{axially symmetric top}). As we repeat the map $n$ times we get
\begin{gather}
X_1^{(n)}=x_1,\nonumber\\
X_2^{(n)}= x_2 \cos(n\Omega)+x_3\sin(n\Omega), \label{rotation}\\
X_3^{(n)}= x_3\cos(n\Omega)-x_2\sin(n\Omega) .\nonumber
\end{gather}

The periodicity condition of period $n$ in this map can be read of\/f directly from 
(\ref{rotation}) as $\cos(n\Omega)=1$, or $\Omega={2\pi\over n}$. 
This condition f\/ixes the values of $x_1$ for each period, according to the rule
\begin{gather}
x_1=\pm\mu_n{2I_2\over I_2-I_1},\qquad \mu_n=\sqrt{{1-\cos(2\pi/n)\over 1+\cos(2\pi/n)}},\qquad n=1,2,3,\dots .
\label{x_1=}
\end{gather}
For small $n$'s we have
\[
\mu_1=0,\quad \mu_3=\sqrt 3,\quad \mu_4=1,\quad \mu_5=\sqrt{5-2\sqrt 5},\quad \mu_6=2-\sqrt 3,\quad \dots.
\]

The conditions determine planes which are orthogonal to the axis of symmetry and intersect
 the axis at certain points def\/ined by (\ref{x_1=}), dif\/ferent for each period. These planes
\begin{gather}
v^{(n)}_{\rm axial\ symm}=\left\{{\mathbf x}\, \big|\, x_1^2=\mu_n^2{4I_2^2\over (I_1-I_2)^2}\right\},
\qquad n=2,3,4,\dots 
\label{axially symm ivpp}
\end{gather}
are the invariant varieties of periodic points characterized by the relations among the invariants:
\[
I_2H_1-H_2={\mu_n^2\over 1+\mu_n^2}{4I_1I_2^2\over I_2-I_1}.
\]

In terms of the variables $(\xi_1,\xi_2,\xi_3)$, the conditions $I_2=I_3$ 
and (\ref{x_1=}) are equivalent to $(\xi_1,\xi_2,\xi_3)=(-\mu_n^2,0,0)$. 
We notice that $v^{(3)}_{\rm axial\ symm}$ is a special case of $v^{(3)}$ in (\ref{rururu}). 
When $n=4$ there is no invariant variety for generic values of $(I_1,I_2,I_3)$, hence 
we are not able to derive $v^{(4)}_{\rm axial\ symm}$ as a special case. We notice that 
the periodicity conditions $(\xi_1,\xi_2,\xi_3)=(-1,0,0)$ in the case of period 4 
are compatible with the conditions $\gamma^{(4)}_2=\gamma^{(4)}_3=0$ of (\ref{gamma4xi}). 

The meaning of the planes presented in (\ref{axially symm ivpp})
 is quite interesting. Because of the symmetry the angular velocity 
 along the symmetry axis is constant as it is expected naturally. An 
 interesting feature is that the value of this angular velocity $x_1$ 
 is `quantized' to some specif\/ic values determined by the shape of the 
 top and dif\/ferent for each period, such that the `angular velocity'~$\Omega$ 
 is quantized to $2\pi/n$ to generate the periodic maps. This is true irrespective 
 to the values of other angular velocities $x_2$ and $x_3$. The generation of the invariant 
 varieties $v^{(n)}_{\rm axial\ symm}$ follows to this fact.

\LastPageEnding
\end{document}